\documentclass[11pt]{article}
\input{psfig}

\begin{document}



\newcommand{\hcm}[1]{\makebox[#1cm]{}}
\newcommand{\Vpd}{V_{pd}}
\newcommand{\Vpp}{V_{pp}}
\newcommand{\Dpd}{\Delta_{pd}}
\newcommand{\pd}{{\it p-d }}
\newcommand{\cuo}{ {CuO$_2$} }
\newcommand{\ppp}{ ppp OVDJE SMO STALI !}
\newcommand{\pll}{\parallel}
\newcommand{\veps}{\varepsilon}
\newcommand{\kk}{{\vec{k}}}
\newcommand{\ide}{\rightarrow}
\newcommand{\RR}{{\vec{R}}}
\renewcommand{\aa}{{\vec{a}}}


\noindent
32$^{nd}$ Winter School of Theoretical Physics,\\
Karpatz, Poland, February 19-29, 1996 \\ 
{\it Solid State Physics:
 From Quantum Mechanics to Technology}\\
(to be published in Lecture Notes in Physics, Springer 1996)

\vskip 1.5cm


\begin{center}
{\large CHARGE DYNAMICS IN CUPRATE SUPERCONDUCTORS} 
\\ [1cm]
E. Tuti\v{s} \\ [2mm]
Institute of Physics of the University of Zagreb, \\
P.O.B. 304, Zagreb, Croatia  \\ [4mm]
%
%
H. Nik\v{s}i\'{c} and S. Bari\v{s}i\'{c}\\ [2mm]
Department of Physics, Faculty of Sciences, \\
P.O.B. 162, Zagreb, Croatia \\ [1cm]
\end{center}

\vskip 1cm


\begin{center}
\hyphenpenalty=10000
\exhyphenpenalty=10000
\sloppy
\parbox[t]{12cm}{
In this lecture we present some interesting issues that
arise when the  dynamics of the charge carriers in the
CuO$_2$ planes of the high  temperature superconductors is
considered. Based on the qualitative  picture of doping,
set by experiments and some previous calculations, we
consider the strength of various inter and intra-cell
charge transfer  susceptibilities, the question of Coulomb 
screening and charge collective modes.  
The starting point is the usual {\it p-d} model extended 
by the long range Coulomb (LRC) interaction. 
Within this model it is possible to examine the 
case in which
the LRC forces frustrate the electronic
phase separation, the  instability which is present in 
the model
without an LRC interaction.   While the static dielectric 
function in such systems is negative  down to arbitrarily
small  wavevectors, the system is not unstable.  We
consider the dominant electronic  charge susceptibilities
and possible  consequences for the lattice properties.}
\end{center}


\vskip 1cm 

\section{Introduction}

A decade of experimental and theoretical research indeed showed 
that the physics of the high temperature copper-oxide
superconductors  is both complex and intriguing.  One major
source of complexity lies in the fact that the electronic 
correlations in these materials are strong.  The
antiferromagnetism in undoped LaCuO$_4$ and YBa$_2$Cu$_3$O$_6$
is  clear evidence of that.  A second source may be found in
experiments that suggest that the  electron-phonon interaction
may be strong as well.  As far as the effects of strong
electron-electron interactions on the  electronic properties of
doped superconducting materials are concerned,  two different
viewpoints exist in general.  From the first point of view, the 
spin correlations, although much  weakened on doping, still
dominantly determine the carrier dynamics.   From the second
viewpoint the superconducting cuprates are doped  charge transfer
insulators, in which the tendency towards spin  ordering  may be
an important, but still secondary, effect.  This second approach
focuses on the charge degrees of freedom  and charge
fluctuations, and the effects that they may have on 
superconductivity \cite{VEXC,LITJR}, electronic properties in
the  metallic phase, and the  crystal lattice  dynamics
\cite{BSEP}.  Some of these issues are the subject of this
lecture.  We consider dominant static charge susceptibilities, 
the collective modes and the dielectric properties   for the
three dimensional system of  copper-oxide planes.  The features
that emerge are related  to the properties of these materials
observed in experiments.  In particular, these are the anomalies
found in the modes of  oscillations of in-plane oxygen atoms, 
the apex oxygen position and movement and the incommensurate and
commensurate  deformations in the direction  parallel to planes. 
In this lecture we consider the system that is close  to the
doped charge transfer insulator.  However, some of the
conclusions remain qualitatively valid away  from this regime.
This is in particular true for the foregoing discussion  related
to the oxygen-oxygen charge transfer modes and the discussion 
regarding the frustrated  phase separated instability.

\section{The p-d model with long range Coulomb interaction}

The presence of \cuo planes is the mayor feature of all  cuprate
high temperature superconductors. The  electronic hopping 
between the planes is relatively small. The `insulating' layer
between the planes  may be composed by atoms  of various
elements.  Another generic  feature, as shown by many
experiments  (e.g., EELS, X ray absorption, photoemission),   is
that on doping the excess holes mostly populate the 
$p_{x,y}-\sigma$ orbitals of the  in-plane oxygen atoms.    These
orbitals further hybridize with some, mainly    $d_{x^2-y^2}$,
orbitals on the copper atoms.  Together with the fact that
correlation effects are important  for the in-plane dynamics of
electrons, this constitutes the basis  of the \pd model. 

The original \pd model \cite{EMERY87,VEXC} consists of the 
tight-binding  part and the part accounting for the short range
Coulomb interaction,
\begin{equation}
H_{pd}=H_{0tba}+H_{SRC}.
\end {equation}
In the tight-binding part the orbital energy levels 
($\veps_d$, $\veps_p$, $\Dpd\equiv\veps_p-\veps_d$) 
are specified and various hybridization terms 
(copper-oxygen, $t_0$, oxygen-oxygen, $t'$, etc.) included. 
$H_{SRC}$ contains the terms describing the Coulomb  
repulsion on the copper, $U_d$, and oxygen site, $U_p$, 
as well as the terms describing the interaction 
between electrons on neighbouring sites (copper-oxygen $\Vpd$, 
the  oxygen-oxygen $\Vpp$, etc.)

In this lecture we will consider the extension of this model
which includes the long range Coulomb forces. This extension is
necessary in order to examine the capability of strongly
correlated electrons  to participate in the screening and to
determine  the effects that strong local forces have on 
electronic plasmon.  The second reason for the inclusion of the
long range  Coulomb forces is to stabilise the system against the
phase separation,  the instability that occurs in the part of the
parameter  space of the original \pd model.

The introduction of long range  Coulomb forces into  the \pd
model is relatively  straightforward.  However, a few points
should be emphasized.  First,  as soon as long range Coulomb
forces are considered,  it is natural to extend the model from 
one plane to  the  three dimensional collection of parallel
planes.  The electron-electron interaction is 
described\footnote{As in the original \pd model,  we neglect the 
Fock terms here.  Physically, this is  correct for the long
range,   but not necessary correct for the short range  Coulomb
forces.} 
by
\begin{equation}
H_{coul} = \frac{1}{2} \sum_{\RR,\RR^\prime,\alpha,\alpha^\prime}
n_\alpha(\RR)
V_{\alpha\alpha\prime}(\RR+\aa_{\alpha}-\RR^\prime-\aa_{\alpha^\prime})
n_\beta(\RR^\prime)
\end {equation}
where $\RR=(R_\pll,\RR_\perp)$ represents the cell index 
and $\alpha$ denotes the orbital of the atom positioned  
at $\aa_\alpha$  with respect to the origin of the unit cell.
The potential $V_{\alpha\beta}(\RR+\aa_{\alpha}-\aa_{\beta})$,
describes the interaction of electrons on different atoms
in the unit cells separated by $\RR$. It is proportional to 
$1/\left|\RR+\aa_{\alpha}-\aa_{\beta}\right|$ at long distances
and reflects the relative position and  the charge distribution 
in the orbitals at short 
distances. The Fourier transform, 
$
V_{\alpha,\beta}(\kk) =
\sum_\RR V_{\alpha\beta}(\RR+\aa_\alpha-\aa_\beta)\exp(i\kk\RR) 
\equiv \tilde{V}_{\alpha\beta}(\kk) 
\exp(-i\kk\aa_{\beta}-i\kk\aa_{\alpha})
$ 
behaves like $1/\kk^2$ at long wavelengths, while
the short wave dependence reflects the structure and 
strength of the local forces.
This may be emphasized by writing $\tilde{V}_{\alpha\beta}(\kk)$  
in the form\footnote{A somewhat more complicated, 
but less suggestive form should be preferable for large $\kk$, 
close to the zone 
boundary. For example, an additional $\exp(-\kk^2a^2)$ factor 
may be introduced in order to make the cutoff at the 
zone boundary soft. Also, it is possible to replace $\kk^2$ by 
$\sum_{l=x,y,z}\left[2-\cos(k_l d_l)\right]/d_l^2$ in order to 
have zero derivatives at the zone boundaries.
} 
\begin{equation}
\tilde{V}_{\alpha,\beta}(\kk) = 
\frac{4\pi e^2}{( a^2d_\perp \veps_\infty) \kk^2}
+ C_{\alpha\beta}(\kk)
\end{equation}
where $a^2d_\perp$ is the volume of the unit cell ($a$ is the 
lattice constant of the \cuo plane and $d_\perp$ is the distance 
between neighboring planes).   The high frequency atomic
screening is included through $\veps_\infty$.  The terms
$C_{\alpha\beta}(\kk)$ have finite  values\footnote{It is an
instructive exercise to reformulate  the calculation of the
Madelung energy of some pure ionic crystal  after the outlined
procedure.  The contribution of the $1/\kk^2$ term then
explicitly  cancels out due to the neutrality of the crystal,
$\sum_\alpha n_\alpha=0$, and the Madelung energy per unit cell
acquires the form  $E_M/N=(1/2) \sum_{\alpha,\beta}
C_{\alpha\beta}(0) n_\alpha n_\beta $.} at $\kk=0$.

As usual, the $\kk=0$ term of the $1/\kk^2$ part of the 
Hamiltonian cancels out because of the neutrality of the
crystal.  Further on, the `passive' orbitals on the ions with
fixed valence  may be excluded from the electronic Hamiltonian,
leaving only the  active orbitals in the copper oxide planes in
the model.\footnote{The terms neglected should be kept in mind as
the  source of a shift, presumably small, of the  energy levels
of active  orbitals with doping. More important, these terms
reappear as the  electron-phonon coupling terms when lattice
movement is considered.}

The Coulomb part of the \pd model, with the long range forces 
included, finally reads as 
\begin{equation}
H_{coul} = \sum_{\kk\in B.Z.,\alpha,\beta}\frac{1}{2}
\left[
(1-\delta_{\kk,0})\frac{4\pi w t_0}{a^3\kk^2} +
C_{\alpha\beta}(\kk)
\right]\tilde{n}_\alpha(\kk)\tilde{n}_\beta(-\kk)
\end {equation}
where $\tilde{n}_\alpha(\kk)
\propto\sum_\RR n_\alpha(\RR)\exp(-i\kk\RR-i\kk\aa_\alpha)$
and the dimensionless parameter 
$w=e^2/d_\perp t_0 \veps_\infty$, measuring the strength of the 
long range Coulomb interaction is introduced. 
The approximate value for $w$ expected for cuprate superconductors 
may be estimated by taking 
$d_\perp\sim 12$\AA, $\veps_\infty\sim3$, $t_0\approx 1.3$eV. 
This gives $w\sim 0.25$.

In the text that follows we will consider the situations with 
and without the $1/\kk^2$ term in the Hamiltonian.  In the latter
case, $C_{\alpha\beta}(\kk)$  corresponds  to  $U_d$, $\Vpd$,
$\Vpp$ usually used in the literature. Physically, they represent
the effective,  screened electron-electron interaction.  In the
case where $1/\kk^2$ is explicitly included  we continue to use
the notation $U_d$, $\Vpd$, $\Vpp$, etc.,  although the physical
meaning of $C_{\alpha\beta}$'s  in that case is somewhat
different. They represent the  local field corrections to the 
$1/\kk^2$ interaction,  presumably large for short distances.

\section{Qualitative features of the electronic spectrum of doped charge 
transfer insulator}

Various estimates of the local Coulomb terms in the \pd model 
emphasize the large repulsion on the copper site, $U_d\sim
10$~eV.   The limit $U_d\ide\infty$ is frequently considered in
the literature,  and we will work in that limit as well.  The
$n_d\leq 1$ constraint forced by infinite $U_d$ is often  dealt
with by introducing the auxiliary (slave) boson field $b$ and the
auxiliary   field fermion $f$ in terms of which the real electron
operator on the copper site becomes  the composite object,
$d_\RR=b_\RR^\dagger f_\RR$. The next step is usually  to
consider the mean field approximation  for the boson  field (the
{\it saddle point approximation} in the  partition function/path
integral language).  Usually, the corrections beyond the mean
field  (the Gaussian fluctuations around the saddle point) are
also considered.  While rigorous justification for this procedure
is usually  sought in the $1/N$ expansion of  the generalised 
model (i.e. N spin components instead of two),  the physical
appeal  comes from the fact that it grasps some major physical
features.  In particular, this applies for related  models
\cite{PCAM}, that may be  more accurately  investigated by other
means (for instance, the  appearance of the Abrikosov-Suhl 
resonance in the Anderson model).  For the \pd model this
procedure  qualitatively reproduces some basic  experimental
facts in cuprates as are the appearance of the in-gap resonance
\cite{INGAPEXP}  and  the large Fermi surface \cite{FSARPES} on
doping. A pedagogical survey of the method and some of its 
results may be  found in ref. \cite{GKJR}.

What basically happens at the mean field level \cite{KLR} for
the slave boson  in the  \pd model  with large $U_d$ is: a) the
boson field $b$ acquires a static component  $B_0$ (with
$\bar{n_d}=1-|B_0|^2 \leq 1$) and an oscillatory  component at
the frequency $\lambda$;  b) $\lambda$ also represents the shift
of the copper  orbital energy on changing from original electron 
to auxiliary fermion fields, $\veps_f=\veps_d+\lambda$;  the
copper-oxygen hybridization in the auxiliary  fermion Hamiltonian
is reduced with respect to the original  hybridization,
$t=t_0B_0$.

\begin{figure}[ht]
\centerline{
\psfig{figure=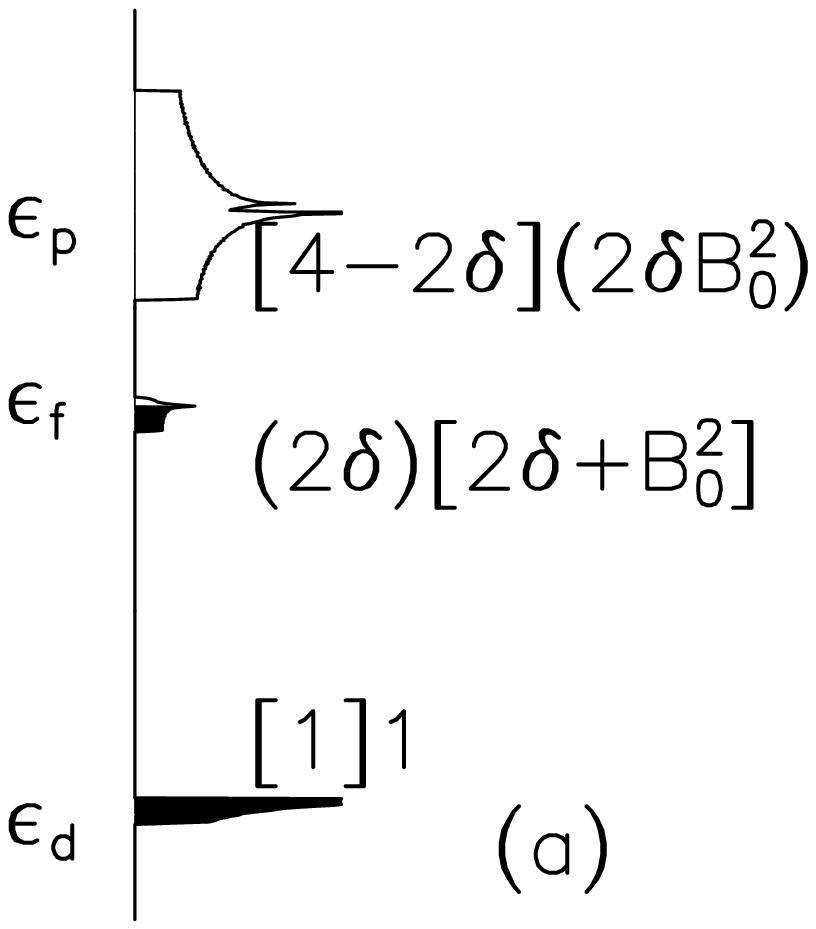,width=3.5cm,height=4cm}\hcm{1}
\hcm{0.5} \psfig{figure=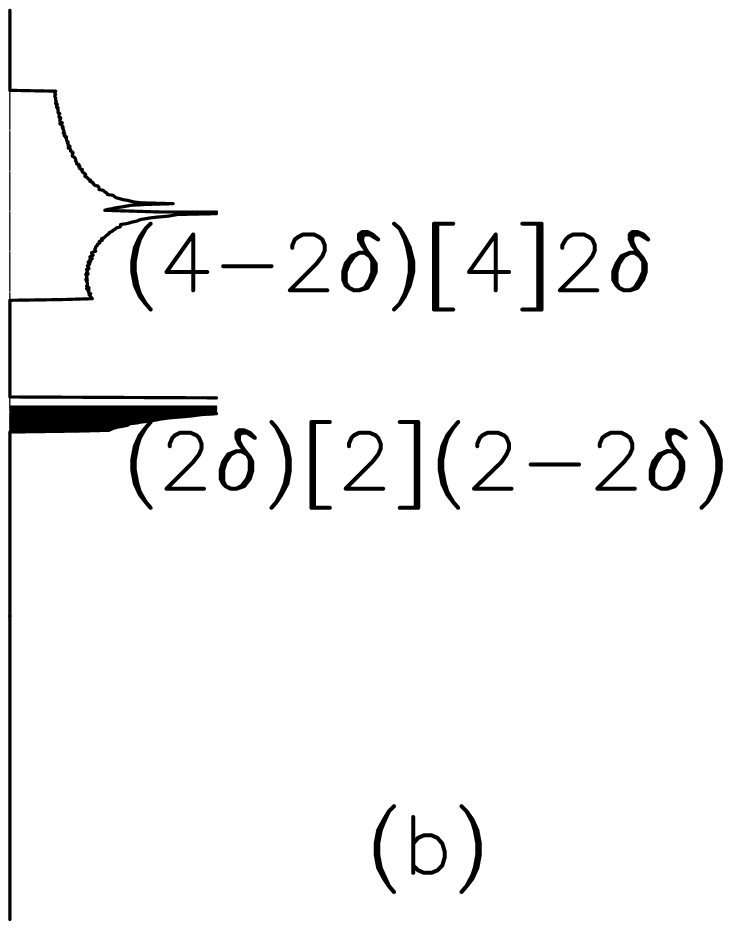,width=3.5cm,height=4cm} 
}
\caption{\small a) The density of states in doped \pd system, as
obtained at the mean field level. b)  The density of states for 
the auxiliary fermions. The labels \protect{$(p)[t](d)$}  
indicate the total weight of the particular part  of the spectrum
\protect{$[t]$} and the shares  \protect{$(p)$} and
\protect{$(d)$} of the oxygen and the copper orbitals,
respectively. The resonance at \protect{$\veps\approx\veps_f$} in
(a) and the corresponding  band in (b) are approximately half
filed for small doping level  \protect{$\delta$}.   of  the
oxygen \protect{$(p)$} and the copper  \protect{$(d)$}. } 
\label{fgGRDOS}
\end{figure}

The band that occurs at $\veps\approx\veps_f$ ($f$-band) in the
auxiliary fermion density of states becomes the  in-gap resonance
when the real electron spectrum is considered. Quite simply, the
majority of the spectral weight for  the copper site that lies in
the auxiliary fermion  $f$-band is transfered back
\cite{ETTZ,PCAM}  to original energy $\veps_d=\veps_f-\lambda$ 
on calculating the auxiliary fermion - auxiliary  boson
convolution in the real electron Green function 
$(-i)\left\langle T (b^\dagger f_\sigma)(t)  (f^\dagger
b)(0)\right\rangle$.  The spectra are illustrated in Fig.
\ref{fgGRDOS}.

The Fermi energy stays in the resonance. The Fermi surface
states  are predominantly made of oxygen orbitals, especially in
the  doped charge transfer insulator (CTI) regime
($\Delta_{pd}\gg t_0$). In that regime the mean field value $B_0$
vanishes when  the doping $\delta$ (measuring the hole
concentration with  respect to the level of one hole per cell)
goes to zero,  $B_0^2\propto \delta$. The renormalized auxiliary
fermion  bandwidth  $\tilde{W}\propto B_0^2$,  being also the
effective width of  the resonance, and the  number of states in
the resonance are also proportional to  doping level $\delta$.

It will be useful to keep this qualitative  
picture\footnote{This picture  may be refined by calculating the 
auxiliary field propagators beyond the saddle point  level
\cite{PCAM,GKJR,NTBR} and by calculating the electronic  spectrum
beyond the decoupling approximation, leading to simple 
convolution.} in mind when the results for the charge
susceptibilities   are discussed.

\section{Charge transfer susceptibilities and Coulomb \\ screening}

\begin{figure}[ht]
\centerline{\psfig{figure=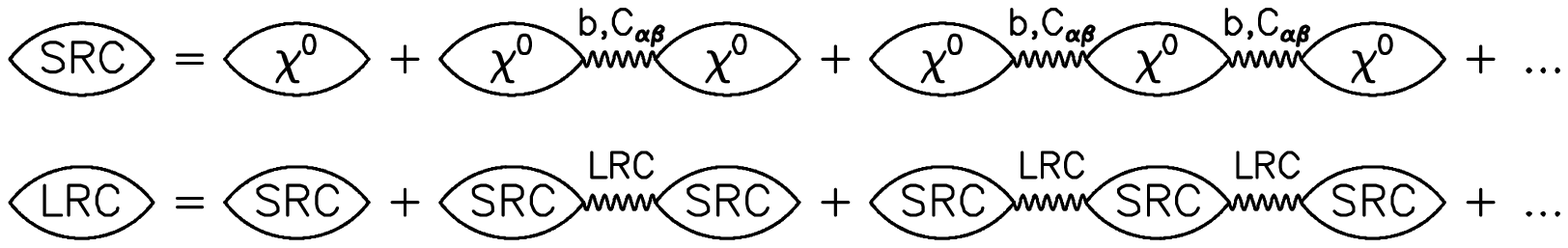,width=10cm,height=1.6cm}}
\caption{\small The corrections beyond the saddle point and Hartree
approximations. }
\label{fgGENRPA}
\end{figure}

The results for the charge susceptibilities that we present are 
calculated on taking into account the first fluctuation
correction  beyond the mean field approximation.  While for large
$U_d$ the slave boson approach is used, for other,  smaller
Coulomb terms we use the  Hartree\footnote{The Hartree
approximation and the RPA fluctuation  correction does not seem
to provide an adequate treatment for  the short range Coulomb
interaction terms as is the $\Vpd$ term.  However, we will really
consider only some qualitative features  that these interaction
bring in. These features are expected  to remain present even if
some more adequate treatment like  the full Hartree-Fock
approximation is used.} approximation \cite{GKJR,GRPSI} as the
mean field  approximation. The fluctuation correction for the
long range Coulomb  interaction corresponds to the usual random
phase approximation (RPA). Fig. \ref{fgGENRPA} reflects the
simple  structure of our calculations.
The charge transfers that we will consider  may be expressed in
terms of the copper and the oxygen  site charges $n_d$, $n_x$,
$n_y$. The combinations reflecting different charge transfer
symmetries and different physics are $n_p=n_x+n_y$ (the charge
on oxygen sites),  $n_{pp}=n_x-n_y$ (the  charge transfer between
inequivalent oxygen sites),   $n_{pd}=n_p-n_d$ (the copper-oxygen
charge transfer) and  $n_{n}=n_p+n_d$ (the total charge in the
cell). The susceptibilities  $\chi_\alpha=\langle\langle
\tilde{n}_\alpha \tilde{n}_\alpha \rangle\rangle$, 
$\alpha=d,p,pp,pd,n$ will be considered.  We will distinguish the
susceptibilities $\chi^0$ of the tight binding fermions with the
mean field  renormalized band parameters; the susceptibilities
$\chi^S$  calculated for the model with short range forces; the
susceptibilities $\chi$   calculated for the model in which  the
long range Coulomb  forces are included as well.  

\begin{figure}[ht]
\centerline{
\psfig{figure=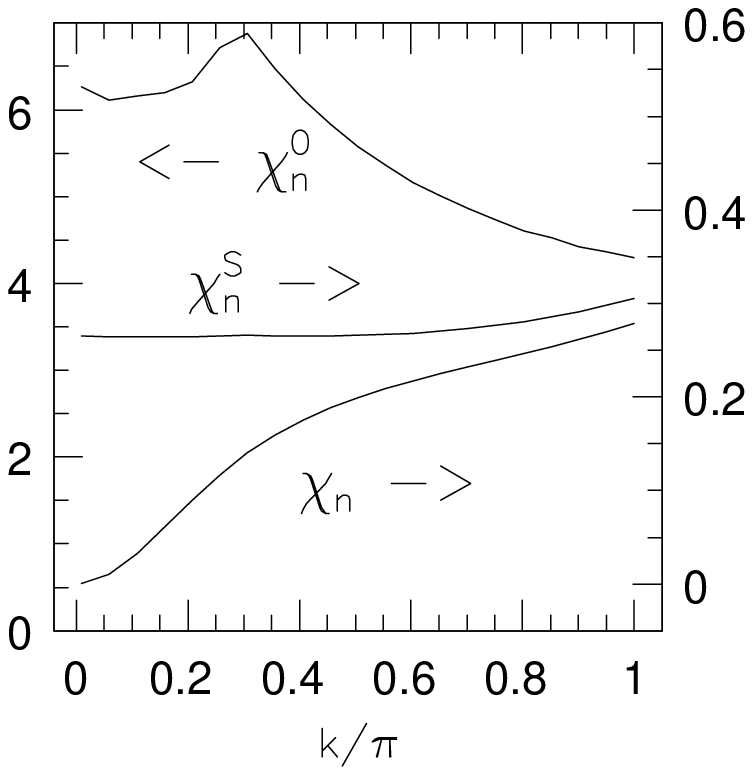,width=5cm,height=5cm}\hcm{0.5}
\psfig{figure=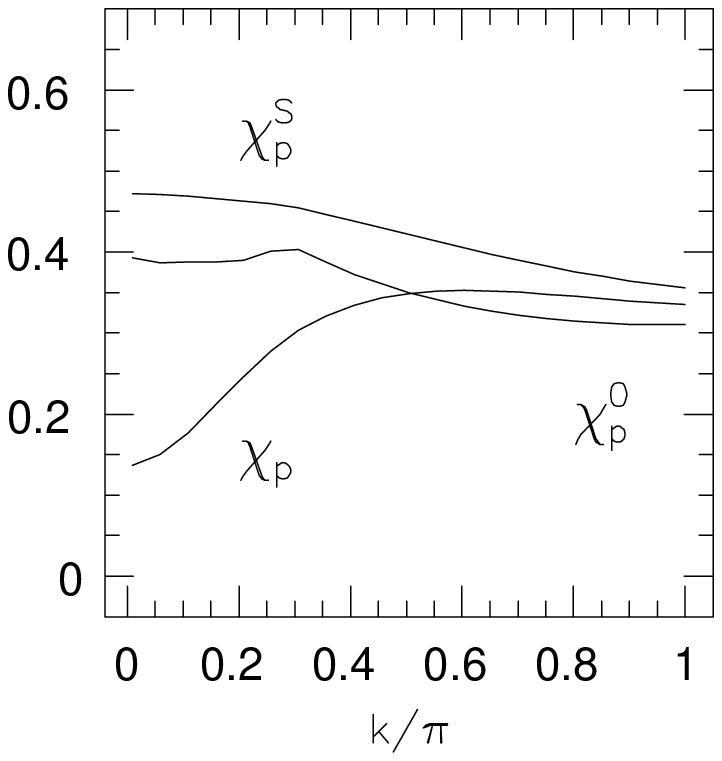,width=5cm,height=5cm} 
}
\centerline{
\psfig{figure=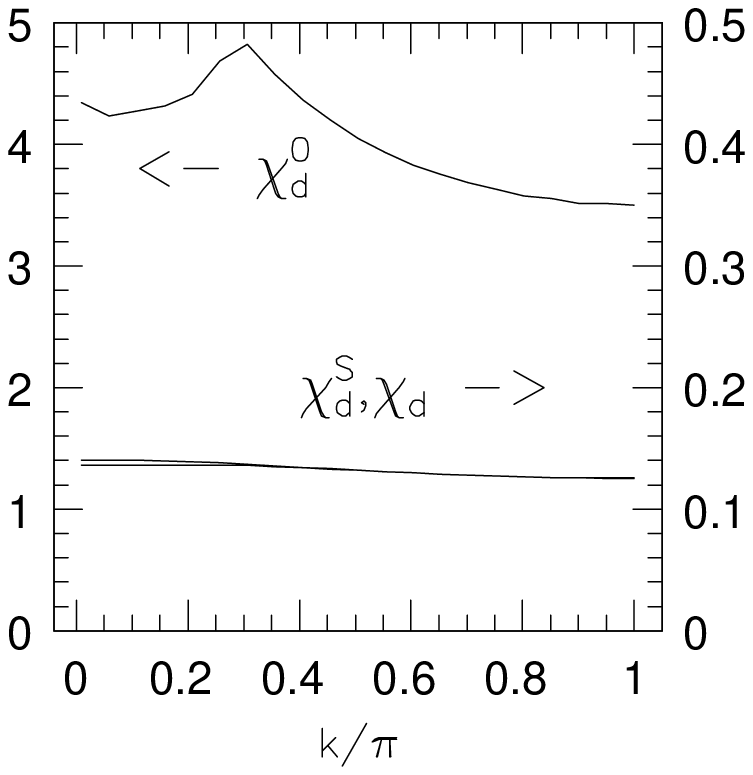,width=5cm,height=5cm}\hcm{0.5}
\psfig{figure=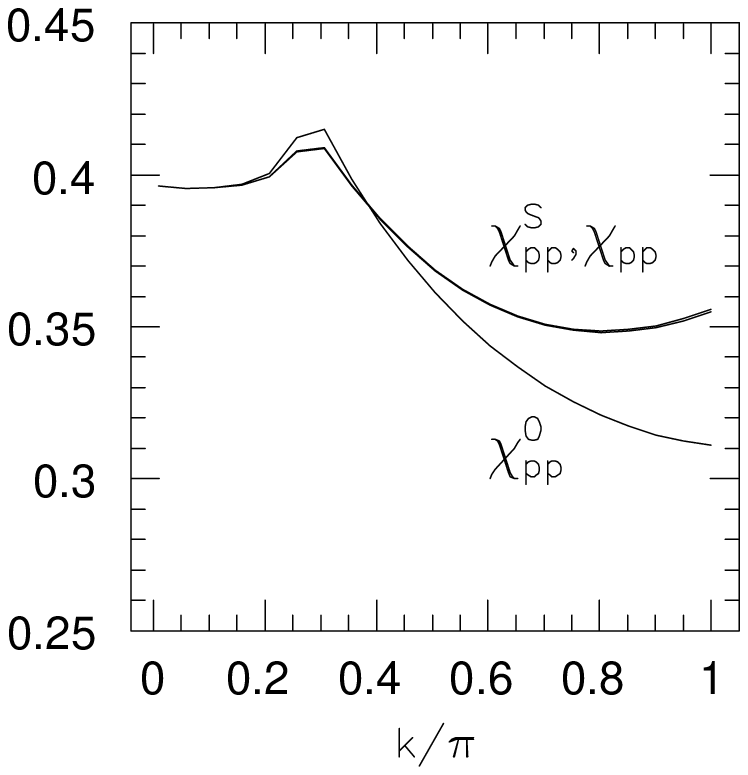,width=5cm,height=5cm} 
}
\caption{\small Static charge susceptibilities  
for \protect{$\Dpd=4,\delta=0.1$} as functions of the wavevector,
\protect{$\kk=(k,0,0)$}: Shown susceptibilities reflect the 
transfers of 
a) the total cell charge b) total oxygen charge in the cell
c) the copper charge d) the intracell oxygen-oxygen 
charge transfer. } 
\label{fgSTchi}
\end{figure}

In Figs. \ref{fgSTchi}  various static charge transfer 
susceptibilities are shown as a function of the wave 
vector $\kk=(k,0,0)$ (hereafter we will use the bare 
copper-oxygen hybridization energy $t_0$ as the energy unit 
and the \cuo plane lattice constant $a$ as the unit distance 
in order to simplify the notation).
From these figures the following may be noted:
\begin{itemize}

\item[a)] The inclusion of the fluctuation of the slave boson
field, corresponding to $U_d=\infty$,  suppresses the charge
fluctuation on the  copper sites for all wavevectors:
$\chi^0_d\gg \chi^S_d$.   The copper-oxygen charge transfer
susceptibility is also diminished. It can be also seen that the
Fermi surface enhancement\footnote{Several types  of the Fermi
surface enhancements  
of the static susceptibilities in the two dimensional  tight
binding model occur at  $\kk\approx 0$ and $\kk\approx
(\pm\pi,\pm\pi)$: van Hove  effects, as well as the nesting or
the sliding Fermi surface  effects, if some flat parts of the
Fermi surface exist.  The degeneracy of these effects for the
square Fermi surface  is lifted for some more  complicated Fermi
surface shapes.}
seen in $\chi^0_d$ at $k_x\approx0.1\pi$ is lost in $\chi^S_d$.

\item[b)] The fluctuation of the  oxygen charge $n_p$ is not 
so much affected by  the auxiliary boson fluctuations.
However, the Fermi surface effect disappears in 
$\chi^S_p$ as well.

\item[c)] The inclusion of the long range Coulomb forces  further
suppresses the total charge fluctuations at small  wavevector,
$\chi_n\ll \chi^S_n$. While this is expected  to happen, it is
important to realize that the charge  fluctuation on the oxygen
sites are mostly affected.  This implies that the holes in the
oxygen  orbitals dominantly participate in the Coulomb screening 
and that their dynamics determines the dielectric properties  of
the system. This issue will become more clear when we  consider
the charge fluctuation spectra. 

\item[d)] The {\it pp} susceptibility related to the charge 
transfer of the between inequivalent oxygen atoms ($x$ and $y$) 
in the unit cell is unaffected both  by the long range Coulomb
interaction and by the  slave  boson fluctuations. The reason
lies in the particular  symmetry of the {\it pp} charge
transfer.  In particular, the Fermi surface effects are 
pronounced in $\chi^S_{pp}$ and $\chi_{pp}$, in contrast  to the
their absence in the susceptibilities $\chi^S_{p}$ and 
$\chi_{p}$ related to the total oxygen charge.  In fact, all van
Hove and nesting enhancements present in the tight binding's
$\chi^0$ remain in $\chi_{pp}$ whereas they are suppressed in all
other channels by the large $U_d$. This, in particular, means
that the $\kk=0$ charge transfer transitions
\cite{BSEP,BSZEL,BSET}  and $\kk=(\pi,\pi)$ charge density wave
should be searched  for mainly in the ${pp}$ channel. The doping
level particularly  favorable is the one for which the Fermi
surface  touches the zone boundary.

\end{itemize} 

\begin{figure}[ht]
\centerline{
\psfig{figure=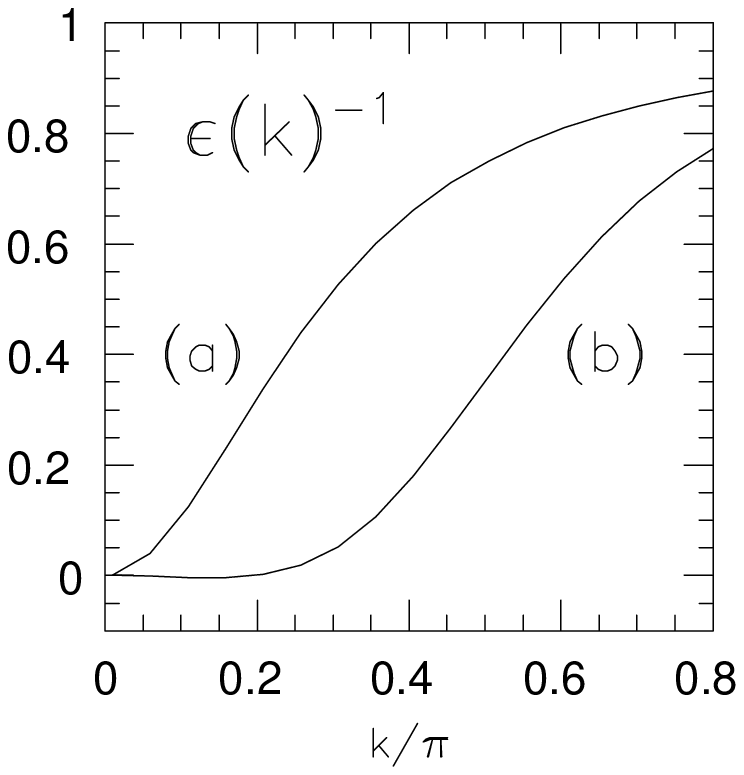,width=5cm,height=5cm}
\hcm{0.5} \psfig{figure=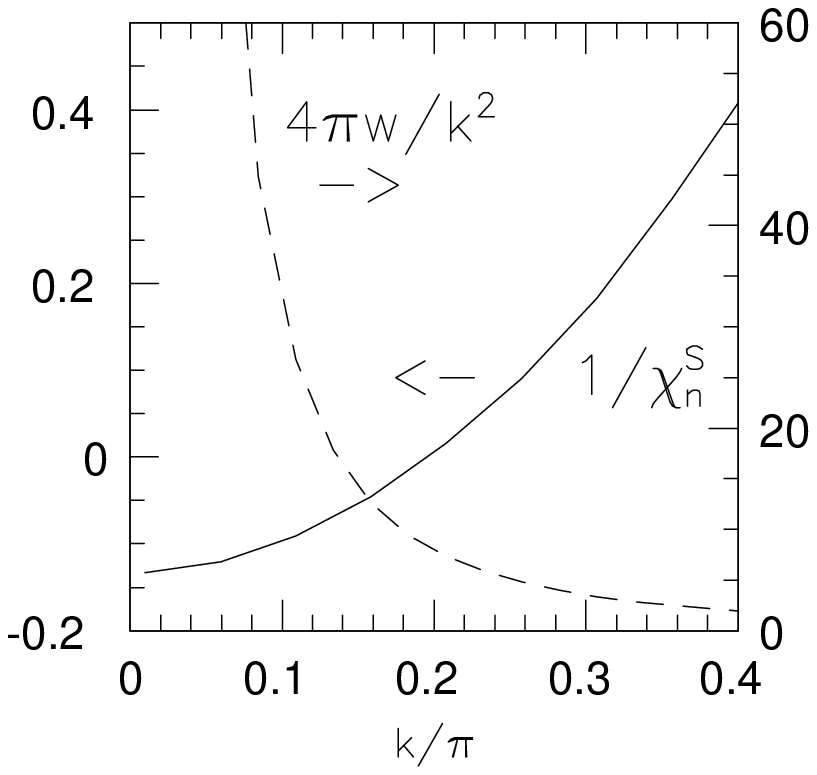,width=5.5cm,height=5cm} 
}
\caption{\small The figure on the left shows 
the dielectric function for 
(a) \protect{$\Vpd=0$} and 
(b) \protect{$\Vpd=1.25$}.  
The second case (as discussed in section 6.) is rather close to the border of 
the `normal' (\protect{$\Vpd<1.15$}) and FEPSI (\protect{$\Vpd>1.15$}) 
situation. The figure on the right shows the total charge 
susceptibility in the unstable \protect{$\Vpd=1.25$} system without the 
long range forces, \protect{$\chi^S_n(0)<0$}.
Other parameters are: \protect{$\Dpd=4$}, 
\protect{$\delta=0.1$}, \protect{$\Vpp=0$}. 
} 
\label{fgepsi}
\end{figure}

The dependence of the static dielectric function on the
wavevector  is shown in Fig. \ref{fgepsi}. The behaviour 
$1/\veps(k)\propto\kk^2\ide 0$ as $\kk\ide 0$ corresponds to the 
full metallic screening. More important, it should be noted that 
the screening distance, expressed as $1/k_{s}$ in Thomas-Fermi 
language,  where $1/\veps(k)=\kk^2/(\kk^2+k^2_{s}$), is of the
order of the lattice spacing, in spite of the fact that  the
concentration of doped holes is small (average distance  between 
them is approximately three lattice constants for  $\delta=0.1$).
The doping dependence of the screening length   shows no drastic
changes with doping. This reflects the fact that  density of
states of the in-gap  resonance (i.e., 
(number-of-states)/(resonance-width)) does not  change very much
with doping, even in the doped CTI regime.

\section{Dynamic correlation functions and collective modes}

\begin{figure}[ht]
\centerline{
\psfig{figure=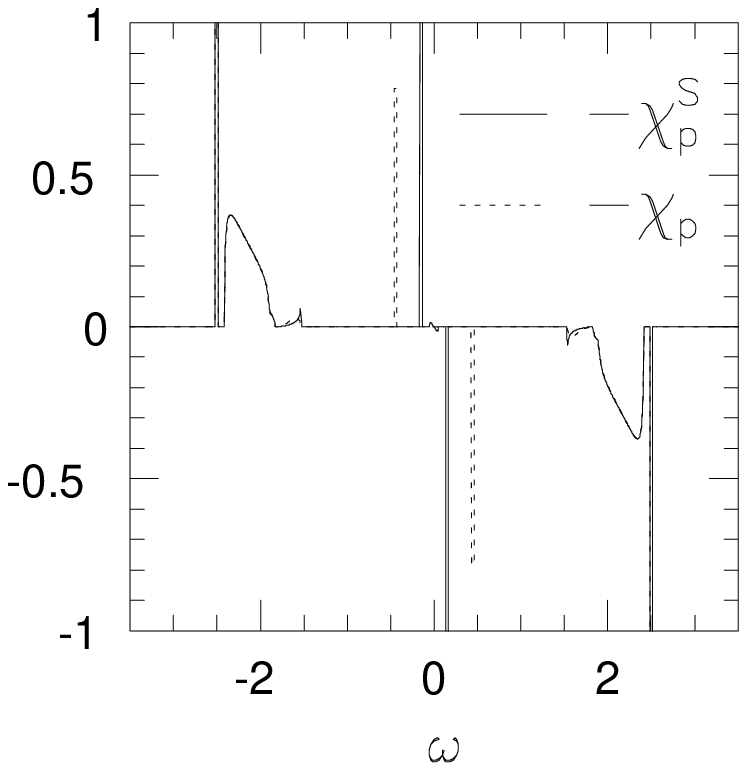,width=5cm,height=5cm}\hcm{2}
\psfig{figure=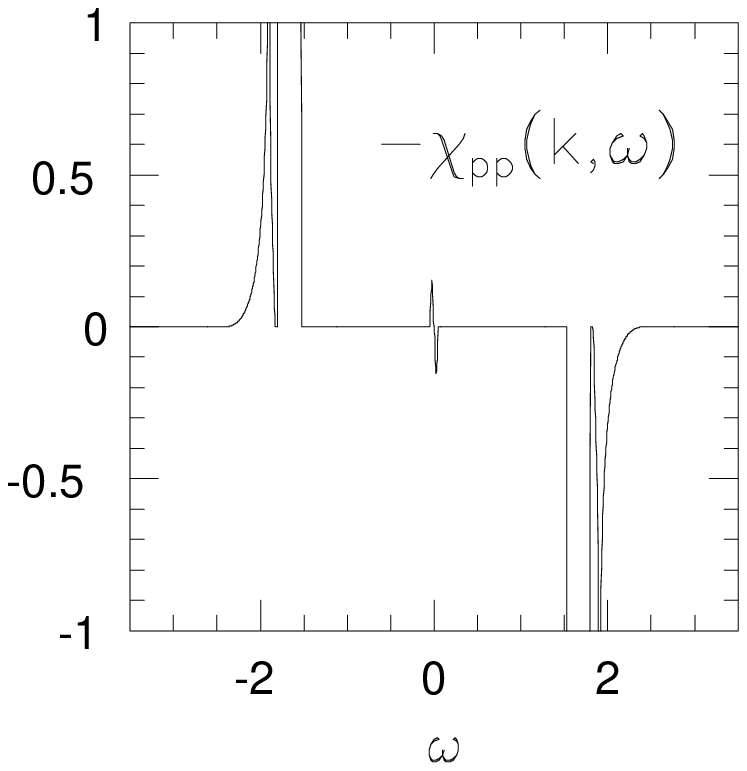,width=5cm,height=5cm} 
}
\caption{\small 
Left: The spectrum of the fluctuation of the charge in the 
oxygen orbitals, \protect{$n_p=n_x+n_y$}. The full line 
correspond to the case without long range interaction. The
inter-cell  and the  intra-cell (Cu-O) charge fluctuations,
corresponding to the zero sound mode and the Cu-O exciton  mode,
and may be distinguished. The  long range Coulomb interaction
changes the zero sound to the intraband plasmon mode, pushing its
frequency to finite frequency at low wavelengths. Right: In the
intra-cell oxygen-oxygen charge  fluctuation spectrum,
\protect{$n_{pp}=n_x-n_y$},   there are no collective modes
present. The strength of the interband contribution in 
\protect{$\chi_{pp}$} is big relative to those in other spectra.
Parameters as before. \protect{$\Vpd=0$},  
\protect{$\kk/\pi=(0.11,0,0)$}. 
  }
\label{fgopsl}
\end{figure}

More detailed information on the charge dynamics are obtained  on
considering the dynamic charge correlation functions.  They
further clarify which orbitals participate in various intra  and
inter-cell charge transfers and how they are affected by short
and long range Coulomb forces. Some aspects of the the spectrum
of charge fluctuations  for the system with large $U_d$ was
considered by Kotliar,  Castellani and coworkers
\cite{GKJR,GREXC}.  They discussed the appearance of the high
frequency  exciton collective mode ($\omega=\omega_{pd}$) as well
as the zero sound mode in the large $U_d$ \pd model.  The
spectrum of the oxygen charge fluctuations    shown in Fig.
\ref{fgopsl} the appearance of these modes in  the spectrum
together with the intraband and interband continua. 
\begin{figure}[ht]
\centerline{
\psfig{figure=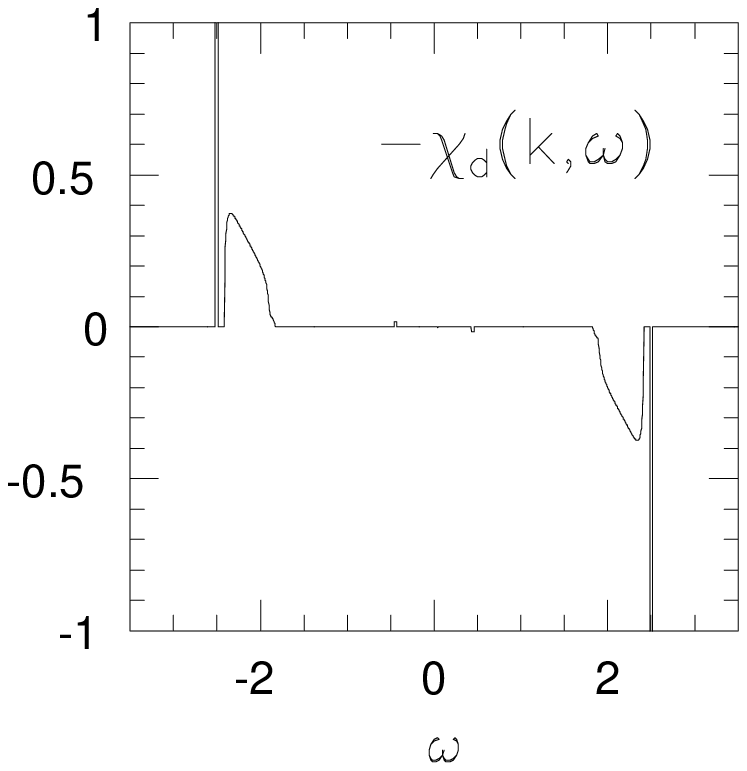,width=5cm,height=5cm}\hcm{2}
\psfig{figure=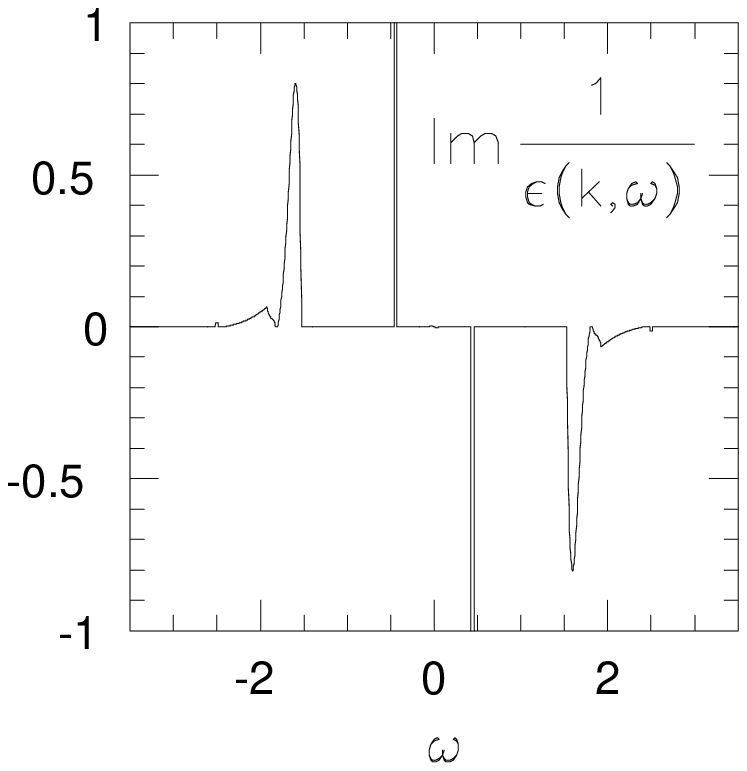,width=5cm,height=5cm} 
}
\caption{\small 
Left: The spectrum of charge fluctuations on the copper site. 
The Cu-O exciton mode and the interband excitations are pronounced.
Right: The total charge fluctuation spectrum as seen through 
       \protect{$1/\veps(\kk,\omega)$}. The intraband plasmon
       and the intra-cell dipolar excitation, which lies within 
       the interband continuum, dominate the spectrum. 
  }
\label{fgodl}
\end{figure}
Fig. \ref{fgopsl}) also shows the effect  of the long range
Coulomb interaction on the spectra. It is interesting to note
that the frequency of the   exciton mode is not at all affected
at long wavelengths. This shows the quadrupolar character of the
corresponding  charge excitations.  On the other hand, zero sound
mode  changes to the intraband plasmon mode with the dispersion 
$\omega_{1}^2(\kk)=\omega_{1p}^2\kk_\pll^2
/(\kk_\pll^2+k_\perp^2$) usual for quasi-2d systems with
negligible  inter-plane hopping.  The frequency (as well as the
strength of the corresponding pole  in the density-density
correlation function) vanishes  for the wave vector directed
perpendicular to the planes. The frequency $\omega_{1p}$ of the 
long wavelength   in-plane oscillations increases on increasing
the strength  $w$ of the long range Coulomb interaction, but
always stays  in the gap. The dependence on $w$ may be understood
through a simple  equation for the Coulomb collective modes in
the two  (infinitely narrow) band systems,
\begin{equation}
1-\frac{\Omega_1^2}{\omega^2}-
\frac{\Omega_2^2}{\omega^2-\Delta_{pf}^2}=0.
\end {equation}
Here $\Delta_{pf}=\veps_f-\veps_p$ measures the renormalised 
gap while $\Omega_1^2$ and $\Omega_2^2$ are products of the 
Coulomb factor $w\propto e^2$ and the intraband 
and the interband oscillator strengths, respectively. 
$\Omega_1$, reflecting the 
intraband charge transfer oscillator strength,
assumes particularly simple doping dependence 
$\Omega_1^2\propto\delta$ in the doped CTI regime. 
This is indeed expected from the picture 
that we have developed until now.

The second solution  of the equation (corresponding to the third 
collective mode in the calculation) represents the intracell 
dipolar mode.  It starts from finite frequency 
$\omega\approx\Delta_{pf}$ for small $w$ and becomes 
the `big plasmon' as $w$ is further increased.
 However, in our calculations the substantial $w>1$ is 
 required to push the frequency of this mode above the 
 interband excitation continuum.

In more detailed inspection of the spectra  $\chi''_\alpha
(\kk,\omega)\equiv {\rm Im} \chi_\alpha  (\kk,\omega)$ of various
charge transfer modes,  we find that the intraband part of the
spectrum is suppressed in all  channels, except from the {\it
pp}  mode.\footnote{None of the collective modes appears
in $\chi''_{pp} (\kk,\omega)$ for $\kk=0$.  For  finite $\kk$
their strength is negligible.}
The zero sound/plasmon mode dominates in the  low frequency part
of spectra of the total charge  fluctuation $\chi''_n
(\kk,\omega)$  and  the oxygen charge fluctuation $ \chi''_p
(\kk,\omega)$.   The strength of this mode in the channels
related to long  wavelength  charge  fluctuations on the copper
site is rather small.  These spectra are dominated by the
interband excitations and the  quadrupolar exciton mode. The
later is absent in the  total charge fluctuations at long
wavelengths. The imaginary part of the inverse dielectric 
function,\footnote{We extract the `macroscopic'  dielectric
function from the matrix of charge  susceptibilities by
considering the coupling of the  system to the external
potential.} is shown in Fig. \ref{fgodl}.

\section{Frustrated electronic phase separation 
instability (FEPSI)}

The dynamics and the dielectric properties of the system with
frustrated electronic phase separation  instability is even more
interesting to consider. Experimentally, the proximity of the
phase separation  instability (PSI) and superconductivity in
cuprates  and related materials is rather well documented
\cite{PSIEXP}. The possibility for  PSI in the model with strong 
electron-electron interaction was addressed by several authors. 
In particular, V.J. Emery and S.A. Kivelson  considered the case
of  low doped antiferromagnet  and discussed the possible
consequences of FEPSI for the  normal state properties and
superconductivity   in cuprates \cite{EKPC}.

The phase separation instability within the \pd model  with short
range forces occurs as the copper-oxygen  Coulomb term $\Vpd$ is
increased in the hole doped system,  as pointed out in ref.
\cite{GRPSI}.  However, the enhancement of the copper-oxygen
charge  transfer by $\Vpd$, often emphasized by some  authors
\cite{VEXC,LITJR}, is not an important  issue here.\footnote{The
value of $\Vpd$ for which the 
enhancement of the copper-oxygen charge  fluctuation takes place
(possibly related to the softening  of the corresponding exciton
mode) is much bigger than  the value required for PSI. For
example, in our calculations for $\Dpd=4t_0$ PSI occurs already 
at $\Vpd=1.15t_0$, while a value approximately two times  bigger
is required for the cooper-oxygen charge transfer  instability to
happen.} \footnote{The \pd model in the parameter range 
$(U_d-\Dpd)\ll\Dpd \ll U_d$  exhibits the phase separation
instability\cite{ETTZ} even for $\Vpd=0$.}
%

The main reason for a system with strong correlations  to choose
the phase separated phase  is to diminish its kinetic energy. The
kinetic energy diminishes on doping since the frustration for the
electron hopping caused by strong  coulomb interaction becomes
less  effective\footnote{V.J. Emery  originally discussed  and
repeatedly emphasized this issues in the  framework low doped
{\it t-J} model.}.  On approaching the point of thermodynamic 
instability the derivative  $(\partial\mu/\partial n) =
(\partial\mu/\partial \delta) $ (the rate of change of the
chemical potential with doping) approaches zero and becomes
negative in the unstable phase. Also, the susceptibility
$\chi^S_n(\kk=0,\omega=0)$ (as well  as all other
$\chi^S_\alpha$'s) diverges at this point, 
$1/\chi^S_n(\kk=0,\omega=0)=(\partial\mu/\partial n)$. An example
of $1/\chi^S_n(k)$ in the unstable system is shown in Fig.
\ref{fgepsi}.

Of course, once the long range Coulomb  interaction is introduced
into the model, the true  phase separated state is not likely to
appear  because of the huge costs in the Coulomb energy.
Speculating on this issue, various authors stop at this  point
and suggest that the charge density wave (CDW) state   will
replace the phase separated state (i.e. that the instability 
will show as the  divergence of $\chi_n$ at some finite
wavevector instead  of at $\kk=0$). However, the pure electronic
CDW (assuming that the lattice is  too rigid to participate in
the formation of the CDW)  corresponds to unrealistically weak
Coulomb forces for the cuprate  superconductors, $w \ll 1$. We
find that the stable FEPSI state with $1/\chi_\alpha(\kk)>0$  and
homogeneous electronic density is more probable. However, the
static dielectric function,
\begin{equation}\label{epsi}
\frac{1}{\veps(\kk)}=\frac{1}{1+(4\pi w/\kk^2) \chi^S_n(\kk_\pll)}
\end {equation}
is negative in FEPSI systems at long wavelengths  since
$\chi^S_n(\kk)$, which accounts for the local  forces, turns
negative,  $1/\chi^S_n(0)\propto(\partial\mu/\partial n)<0$, as
shown  in Fig. \ref{fgepsi}.

The figure suggests the following formula to describe 
the dependence of the dielectric function on the wavevector,  
\begin{equation}\label{gTF}
\frac{1}{\veps(k)}\approx\frac{1}{1-b} 
\left( \frac{k^2}{k^2+k_{s1}^2} - 
\frac{k^2 b}{k^2+k_{s2}^2} \right),
\end {equation}
with $1/k_{s2}\ll k_{s2}$ ($1/k_{s2}$ is at least of the  order
of few lattice constants in the \cuo plane) and $b<1$.  The
formula resembles the Thomas-Fermi formula, but has   two
characteristic wavevectors instead of one.    In order to get a
qualitative feeling for the system, one  may try to play with a
three dimensional toy model with  the dielectric function of the
form (\ref{gTF}). The feature  that readily emerges is the
overscreening of the test charge  at distances $1/k_{s1}$
followed by the complete  screening at distances beyond
$1/k_{s2}$.  This implies the attraction of two equal test
charges down  to distances $1/k_{s2}$.  Also, the limit $1/\veps
\ide 0$ as $k\ide 0$ accounts for the  total metallic screening.
It is easy to see that a piece of the material exposed to the
static  electric field behaves just like an ordinary metal, 
developing a finite surface charge  in order to ensure
$\vec{E}=0$ in the bulk.

\begin{figure}[ht]
\centerline{
\psfig{figure=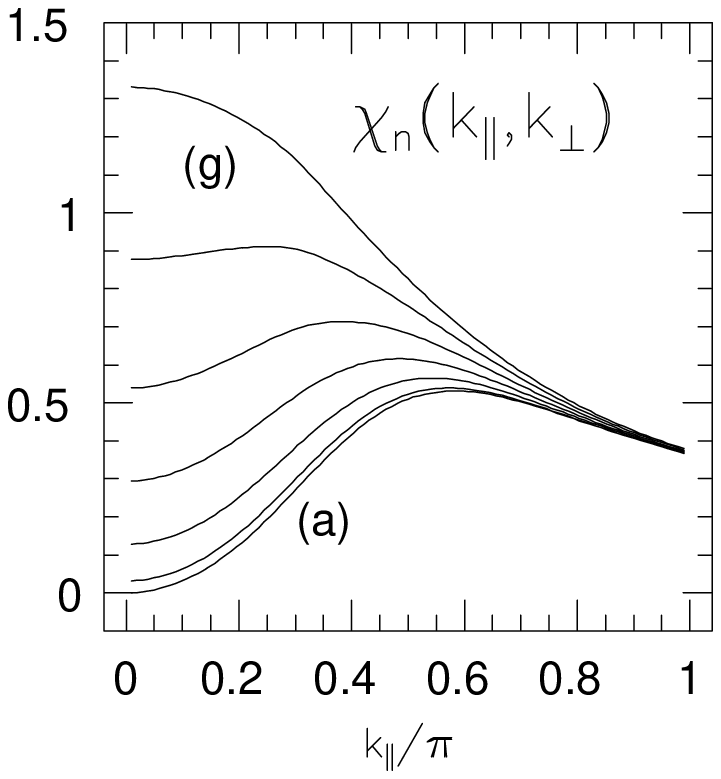,width=5cm,height=5cm}\hcm{2}
\psfig{figure=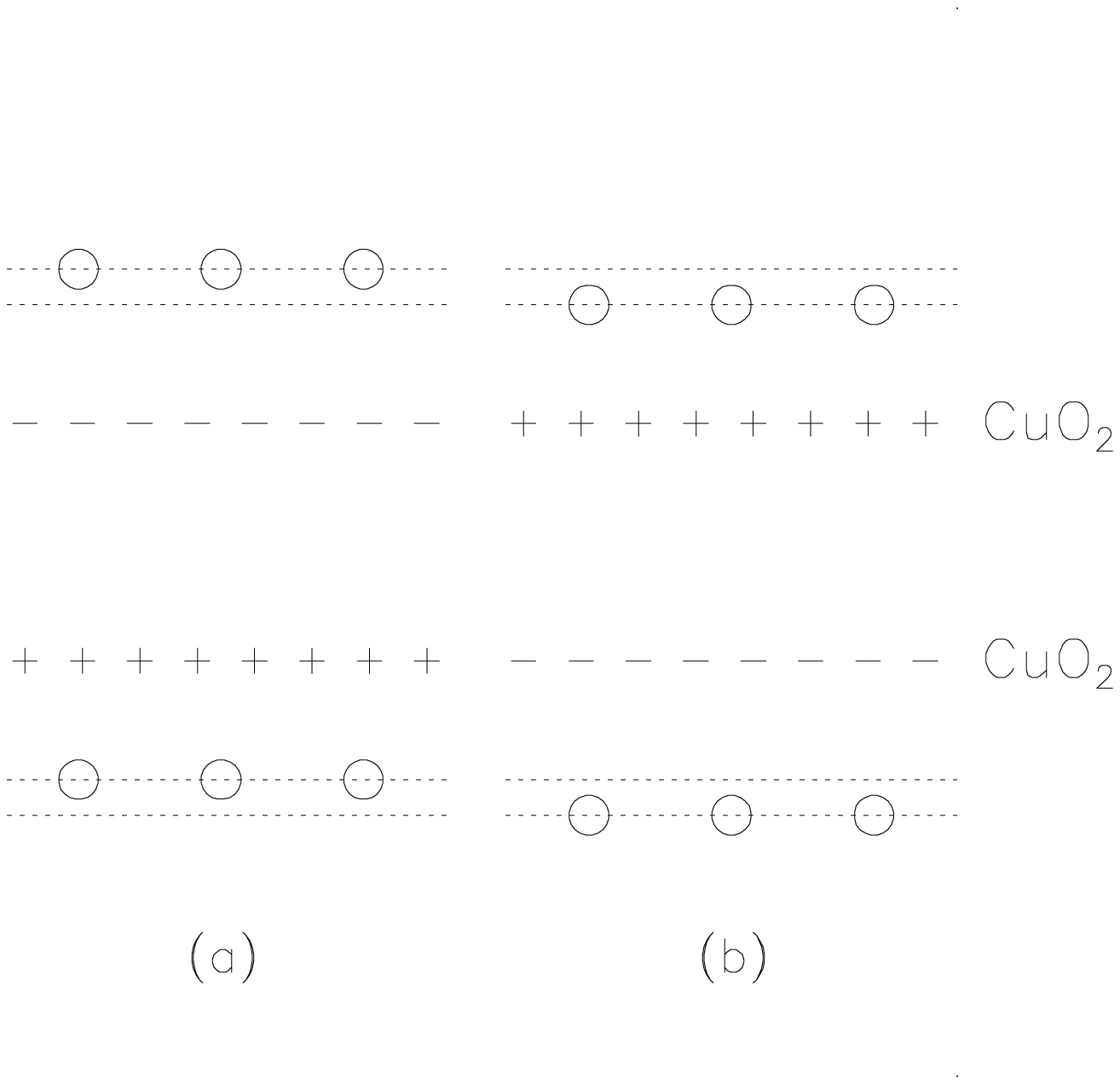,width=5cm,height=5cm} 
}
\caption{\small 
Left: The dependence of the total charge 
      susceptibility on \protect{$k_\perp$}, 
      \protect{$k_\perp/\pi=0(a),0.1,..,0.6(g)$}.
      Right: The possible
     effect of large inter-plane charge transfer 
     susceptibility on apex oxygen atoms in the systems
     with \protect{\cuo} bilayers, like 
     \protect{YBa$_2$Cu$_3$O$_{7-x}$}. 
  }
\label{fgapex}
\end{figure}

Returning to our quasi-two dimensional FEPSI metal we examine its
static charge susceptibility.  It is given in a form that may be
easily analysed   (see again Fig. \ref{fgepsi}), 
\begin{equation}
\frac{1}{\chi_n(\kk)}=
\frac{4\pi w} {k^2_\perp+\kk_\pll^2} + \frac{1}{\chi^S_n(\kk_\pll)}.
\end {equation}
The most pronounced features (see Fig. \ref{fgapex}) are the
maximum at finite $k_\pll=k_c$ (not related to  some particular
Fermi surface wave vector -- $k_c$ decreases with decreasing $w$)
for $k_\perp=0$,  and the increase of $\chi_n$ towards the 
$k_\perp=\pi/d_\perp$ zone boundary. It may be noted here that 
the divergence of $\chi_n(k_\pll=k_c)$ for sufficiently small $w$
is the usually mentioned CDW  instability which comes as the
alternative to the  FEPSI  homogeneous in the \cuo plane.  This
instability corresponds to the softening of  the intraband
plasmon branch at $k_\pll=k_c$.  However, the Cu-O exciton mode
is not affected at this point. 

\section{Lattice properties and the stability of FEPSI systems}

Finally, we may turn to the question of the stability of the
FEPSI system, considering in particular the influence of the
electronic gas  on the ionic lattice (which was considered
infinitely rigid up to now). First, the question of total
compressibility of the FEPSI  system may be addressed. It may be
shown  in a number of ways that the short range forces between
ions  provide the stability of the system. One way is to consider
the  longitudinal sound velocity and the corresponding frequency 
$\omega^2(\kk)=c^2\kk^2=c_O^2\kk^2+\Omega^2_{p,ion}/\veps_{el}(k)$
consisting  of the contribution of the short range forces between
ions and the  long range forces screened by electrons. For FEPSI
systems the second term is negative. In terms of the total bulk
modulus $B$ which determines the sound velocity $c^2=B/\rho$, 
the contribution of the second term   equals $n^2(\partial \mu
/\partial n)<0$.   This, however, does not jeopardize the
stability of the system,  since size of the the electronic
contribution is two orders of magnitude  smaller than the total
bulk modulus measured  for  cuprates,\footnote{The correct order
of magnitude  for the crystal compressibility may by be obtained
from simple Madelung calculations.} $B\sim 10^{-12}{\rm
dyn/cm}^2$. The reason for small absolute value of the 
electronic contribution may be ultimately   traced back to the
small density of the electronic  gas.{\footnote{In the jellium 
language,
$r_s\sim 8$, if we count one hole per unit cell; counting only 
doped holes we get $r_s\sim 16$ for $\delta=0.1$.  At this point,
it seems interesting to note  that the theoretical search for the
high temperature  superconductivity before the discovery of
cuprate superconductors pointed, indeed,  towards the low density
metals \cite{SSSbook}. One of the basic properties of these
systems is the  appearance of the  negative static dielectric 
function. That negative static dielectric  function is the
feature desirable for the electronic mechanism of 
superconductivity was rather pedagogically explained by 
Littlewood in ref. \cite{LITJR}.  However, as  emphasized by 
Kirzhnits at al. \cite{SSSbook}, this is not the sufficient
condition for the superconductivity to occur - the interaction
between test  charges is not the one that enters  the gap
equation - the local field corrections and correlation effects 
that themselves  lead to negative static dielectric function 
should be taken care of, as well.}

More interesting than the compression of the whole system  are
the phonon modes that couple directly  to the charge fluctuations
that  our model calculations distinguishes as relatively big.  In
summary\footnote{The possibility of soft Cu-O charge transfer
exciton is out of focus of this lecture.  However, the formalism
set here is perfectly suitable for its consideration since, as
already explained, one almost inevitably works `deeply' inside
the FEPSI phase.},  these are:  
a)~the in-cell oxygen-oxygen  charge fluctuations
b)~the $k_\perp=\pi$ inter-plane charge transfer  (this becomes
$k_\perp=0$ when systems with \cuo bilayers  are  considered) 
c)~the incommensurate charge density fluctuations inside the 
\cuo planes.  
These modes may be expected to become soft
$\tilde{\omega}^2_{ph}=\omega^2_{ph,0}-g_{ep}^2\chi_{el}$  when
the electron-phonon coupling $g_{ep}$ is substantial or lattice
instabilities  and/or anharmonicities may result. 

In that respect it is interesting to note that, indeed,  in
several copper oxide superconductors  anomalies  in the lattice
modes coupled to the intracell  oxygen-oxygen  charge transfer
are observed. For example,  in Tl$_2$Ba$_2$CaCu$_2$O$_8$ in the
normal state the dominant mode for the  in-plane oxygen motion is
one that couples to $n_{pp}$. This kind of oscillation becomes
suppressed \cite{TOBI} in the superconducting phase. Similar
competition of the superconductivity and lattice deformation
that  couples to $n_{pp}$ is found \cite{LTT,BSEP} in the  LTT
phase of La$_{2-x}$Ba$_x$CuO$_4$.

The discussion on the experimental evidence for  the in-plane
incommensurate  CDW   in related, but non-superconducting
compounds,  may be found in ref. \cite{EKPC}.

Here we would like to point out one possible sign of the FEPSI
situation in YBa$_2$Cu$_3$O$_{7-x}$, the compound  which contains
two \cuo layers per unit cell.  From our discussion we expect to
find  anomalies  for ion movements that couple to the charge 
transfer between  these layers. The IR-active mode of the apex
oxygen atoms is a natural candidate for such coupling.  The
coupling, being strong enough, results in the situation with  two
equivalent, minimum energy  positions for the apex oxygen  atoms
(see Fig. \ref{fgapex} for illustration). Experimentally, this
picture, supported later on by some other  measurements, first
emerged from the analysis of   the EXAFS \cite{oxIRdw} in
YBa$_2$Cu$_3$O$_{7-x}$

\section{Conclusion}

The strong short range Coulomb forces substantially  complicate
the charge dynamics in doped copper-oxide planes.  It seems
difficult to envisage some effective  band picture which would
simultaneously account for the static ($k_{s}$) and dynamic
($\omega_p$) screening as well as  some Fermi surface  effects
that we pointed out.  Also, the strong Coulomb forces may tend to
cause electronic phase separation, the vicinity of which was
pointed out by several  authors as the possible source of
superconductive pairing and the   anomalous normal state
properties. At this point the  introduction of the long range
Coulomb forces into the model seems  crucial, what we did for the
{\it p-d} model. We showed that, while the long range Coulomb
forces  suppresses the instability,  the negative static
dielectric function and, more important, rather  big total charge
susceptibilities in some parts of the $k$-space remain as the
characteristic signs of FEPSI systems. Some of these signs were
found in cuprate superconductors and related materials. While the
lattice instabilities which may occur in order to exploit the
large electronic  susceptibilities probably do not contribute to
the appearance of  superconductivity, the effective electronic
interaction which  drives FEPSI may be favorable in that sense.

{\bf Acknowledgements.} We gratefully acknowledge the 
discussions and remarks by I.~Batisti\'{c} and  D.~K.~Sunko.

{\small
\setlength{\parskip}{0pt}

}


\begin{thebibliography}{the}

\setlength{\itemsep}{0pt}

\bibitem{VEXC} 
C.M.~Varma, S.~Schmitt-Rink, E.~Abrahams,
Solid State Commun. {\bf 62} (1987)  681;
C.M.~Varma et al.,
Phys. Rev. Lett {\bf 63} (1989) 1996;
C.M.~Varma, Phys. Rev. Lett {\bf 75} (1995) 898;

\bibitem{LITJR} 
P.B. Littlewood in:  V.J. Emery(ed.), 
{\it Correlated electron system}, 
World Scientific 1993., {\it p. 1} 


\bibitem{BSEP} 
S.~Bari\v{s}i\'{c}
in: B.K.~Chakraverty (ed.), 
{\it Critical Trends in High $T_{c}$ Superconductivity},
World Scientific, 1991.



\bibitem{PCAM} 
P.~Coleman,
Phys. Rev. B {\bf  35} (1987) 5072



\bibitem{EMERY87}
V.J.~Emery,
Phys.Rev. Lett.  {\bf  58} (1987)  2794



\bibitem{INGAPEXP} 
N. Nucker et al.,
Phys.Rev. B {\bf 37} (1988)  5158; 
T.~Takahashi et al.,
Nature {\bf 334}  (1988) 691;
T.~Watanabe et al., 
Phys.Rev. B {\bf 44} (1991)  5316

\bibitem{FSARPES} 
J.C.~Compuzano et al., Phys.Rev.Lett. {\bf 64} (1990) 2308;
D.S. Dessau et al., Phys. Rev. Lett. {bf 71} (1993) 2781; 
Aebi P. et al., Phys. Rev. Lett. {\bf 72} (1994) 2757


\bibitem{GKJR} 
G. Kotliar  in: 
V.J. Emery (ed.),  {\it Correlated electron system}, 
World Scientific 1993., {\it p. 118} 


\bibitem{KLR} 
G.~Kotliar, P.A.~Lee, N.~Read,
Physica C {\bf 153-155} (1988) 538


\bibitem{ETTZ} 
E. Tuti\v{s}, Ph.D. thesis, University of Zagreb, 1994.



\bibitem{NTBR} 
H.~Nik\v{s}i\'{c}, E.~Tuti\v{s}, S.~Bari\v{s}i\'{c},
Physica C. {\bf 241} (1995) 247-256


\bibitem{GRPSI} 
M.~Grilli, R.~Raimondi, C.~Castellani,
C.~Di Castro, G.~Kotliar,
Phys.Rev.Lett. {\bf 67} (1991) 259


\bibitem{BSZEL} 
S.~Bari\v si\'c, J.~Zelenko,
Solid State Commun. {\bf 74} (1990) 367


\bibitem{BSET} 
S.~Bari\v{s}i\'{c}, E.~Tuti\v{s},
Solid State Commun.  {\bf 87} (1993) 557


\bibitem{GREXC} 
C. Castellani, G.~Kotliar, R.~Riamondi, 
M.~Grilli, Z.Wang, M. Rozenberg, 
Phys. Rev. Lett {\bf 69} (1992) 2009


\bibitem{PSIEXP} 
K.A. M\"{u}ller and G. Bedenek (eds.), {\it Phase
Separation in Cuprate Superconductors}, 
World Scientific, 1993

\bibitem{EKPC} 
V.J.~Emery and S.A.~Kivelson, 
Physica {\bf C 209} (1993) 567;
{\bf C 235-240} (1994) 189


\bibitem{SSSbook} O.V.~Dolgov, D.A.~Kirzhnits and E.G.~Maksimov; 
V.L.~Ginzburg and D.A.~Kirzhnits in:  V.L.~Gizburg (ed.),
{\it Superconductivity,
Superdiamagnetism, Superfluidity}, Mir, Moscow 1987. 


\bibitem{TOBI}
B.H.~Toby, T.~Egami, J.D.~Jorgensen, M.A.~Subramanian,
Phys.Rev.Lett. {\bf 64} (1990) 2414

\bibitem{LTT} 
M.~Sera, Y.~Ando, K.~Fukuda, M.~Sato, I.~Watanabe,
S.~ Nakshima, K.~Kumgai,
Solid State Commun. {\bf 69} (1990) 851
 



\bibitem{oxIRdw} 
J.~Mustre de Leon et al., Phys. Rev. Lett. {\bf 65}, 
(1990) 1675; Phys. Rev. B {\bf 45}, (1992) 2447



\end{thebibliography}
\end{document}